\documentclass[12pt]{article}
\usepackage[affil-it]{authblk}
\usepackage{graphicx}
\usepackage{hyperref}

\usepackage{amsfonts, amsmath, amsxtra, amssymb, latexsym,url}
\usepackage{makeidx}
\usepackage{multirow}
\usepackage{epsfig,epsf}
\usepackage{graphics}

\usepackage{bm}
\usepackage{appendix}

\def\wh{\widehat}

\def\cov{\hbox{cov}}

\def\btheta{{\boldsymbol \theta}}

\newcommand{\bI}{\mathbf{I}}

\newcommand{\bZ}{\mathbf{Z}}

\newcommand{\bC}{\mathbf{C}}

\newcommand{\bL}{\mathbf{L}}
\newcommand{\bD}{\mathbf{D}}

\newcommand{\bs}{\mathbf{s}}

\newcommand{\bv}{\mathbf{v}}

\newcommand{\EXPT}[1]{\mathbb{E}_t\left(#1\right)}
\newcommand{\EXPA}[1]{\mathbb{E}_a\left(#1\right)}
\newcommand{\mydet}[1]{\vert #1 \vert}

\DeclareMathAlphabet{\mathitbf}{OML}{cmm}{b}{it}
\def\thetab{\bm{\theta}}
\def\LL{\mathcal{L}}

\def\bh{\mathbf{h}}

\def\H{\mathcal{H}}

\graphicspath{ {./images/} }

\usepackage{geometry}
\begin{document}
\title{Identification of unknown parameters and prediction with hierarchical matrices}

\author[1]{Alexander Litvinenko
\thanks{Corresponding author, Email: \texttt{litvinenko@uq.rwth-aachen.de}}}
\affil[1]{RWTH Aachen, Aachen, Germany} 
\author[2]{Ronald Kriemann
\thanks{Email: \texttt{rok@mis.mpg.de}}}
\affil[2]{Max Planck Institute for Mathematics in the Sciences (MiS) in Leipzig}
\author[3,4]{Vladimir Berikov
\thanks{Email: \texttt{berikov@math.nsc.ru}}}
\affil[3]{Sobolev Institute of Mathematics,
Novosibirsk, Russia}
\affil[4]{
Novosibirsk State University, Novosibirsk, Russia
}
\date{\vspace{-5ex}}
\maketitle

\begin{abstract}
Statistical analysis of massive datasets very often implies expensive linear algebra operations with large dense matrices. Typical tasks are an estimation of unknown parameters of the underlying statistical model and prediction of missing values. We developed the $\H$-MLE procedure, which solves these tasks. 
The unknown parameters can be estimated by maximizing the joint Gaussian log-likelihood function, which depends on a covariance matrix. To decrease high computational cost, we approximate the covariance matrix in the hierarchical ($\mathcal{H}$-) matrix format.
The $\mathcal{H}$-matrix technique allows us to work with inhomogeneous covariance matrices and almost arbitrary locations. Especially, $\H$-matrices can be applied in cases when the matrices under consideration are dense and unstructured.

For validation purposes, we implemented three machine learning methods: the k-nearest neighbors (kNN), random forest, and deep neural network. The best results (for the given datasets) were obtained by the kNN method with three or seven neighbors depending on the dataset. The results computed with the $\H$-MLE method were compared with the results obtained by the kNN method. 

The developed $\H$-matrix code and all datasets are freely available online.
\end{abstract}
\newpage
\tableofcontents
\section{Introduction}\label{sec:intro}
The number of measurements that should be statistically analyzed increases from year to year. The involved statistical methods often contain intensive  operations with large dense matrices.
In case these measurements are distributed irregularly across the given domain, the efficient algorithms like the Fast Fourier Transformation and similar are not applicable. Therefore, new efficient methods are needed.

To make these expensive computations possible, we suggest to use the hierarchical matrix ($\mathcal{H}$-matrix) technique \cite{HackHMEng, Part1, MYPHD, khoromskij2009application}. We will demonstrate how to solve statistical inference and prediction tasks appearing very often in spatial statistics. This work is an extension of our previous research \cite{LitvHcov19,LitvHLIBPro20}.
The first novelty is that we simultaneously identify four unknown parameters and not three.
This new parameter is the regularization term - the nugget $\tau^2$. Another novelty is the new research on how well we can make statistical predictions with $\H$-matrix approximations. Finally, we compare our identified parameters and predicted values with the actual results and with results obtained by other methods. We summarize the strong and weak sides of the $\H$-matrix technique.

\textbf{Assumptions.} Let $(\bs_1,\ldots,\bs_n)$ be the set of locations. We model the set of measurements as a realization from a stationary Gaussian spatial random field. 
Specifically, we let $\bZ=\{Z(\bs_1),\ldots,Z(\bs_n)\}^\top$, where $Z(\bs)$ is a Gaussian random field indexed by a spatial
location $\bs \in \mathbb{R}^d$, $d=2,3$. Then, we assume that $\bZ$ has zero mean and a stationary parametric covariance function 
$C(\bh;\btheta)=\cov\{Z(\bs),Z(\bs+\bh)\}$, where $\bh\in\mathbb{R}^d$ is a spatial lag vector and $\btheta\in\mathbb{R}^4$
the unknown parameter vector of interest. Statistical inferences about $\btheta$ are often based on the Gaussian
log-likelihood function:
\begin{equation}
\label{eq:likeli}
\LL(\thetab)=-\frac{n}{2}\log(2\pi) - \frac{1}{2}\log \mydet{\bC(\thetab)}-\frac{1}{2}\bZ^\top \bC(\thetab)^{-1}\bZ,
\end{equation}
where the covariance matrix $\bC(\thetab)$ has entries $C(\bs_i-\bs_j;\btheta)$, $i,j=1,\ldots,n$. The maximum likelihood
estimator of $\btheta$ is the value $\wh \btheta$ that maximizes (\ref{eq:likeli}). When the sample size $n$ is large,
the evaluation of (\ref{eq:likeli}) becomes challenging. Indeed, the storage of the $n$-by-$n$ covariance matrix $\bC$ requires ${\cal O}(n^2)$ units of memory. Computation of the inverse and log-determinant of
$\bC(\thetab)$ cost ${\cal O}(n^3)$ FLOPs. Hence, parallel and scalable methods that can reduce this high cost are needed.

Similar works for the case when measurements are located on a rectangular grid can be resolved via the fast Fourier transformation (FFT) method \cite{WHITTLE54, DAHLHAUS87, guinness2017circulant, stroud2017bayesian, Dietrich2} with the computing cost $\mathcal{O}(n\log n)$. However, the FFT method does not work for data measured at irregularly spaced locations or requires expensive, non-trivial modifications.

Other recent ideas include the low-tensor rank methods \cite{litv17Tensor, nowak2013kriging, Huang2016},
covariance tapering \cite{Furrer2006,Kaufman2008, sang2012full,stein2013statistical}, likelihood approximations \cite{Stein2013,Fuentes:2007}, 
Gaussian Markov random-field approximations \cite{fuglstad2015does},
Vecchia framework \cite{ Vecchia88,Katzfuss17}, the nearest-neighbor Gaussian process models \cite{Datta:Banerjee:Finley:Gelfand:2015}, the low-rank update \cite{SaibabaKitanidis15UQ}, multiresolution Gaussian process models \cite{nychka2015multiresolution}, equivalent kriging \cite{kleiber2015equivalent}, and Bayesian-like approach \cite{hermann2016inverse, Matthies16}. 

An $\H$-matrix approximation of covariance matrices was done in \cite{Li14, saibaba2012application, BallaniKressner, harbrecht2015efficient, ambikasaran2013large, BoermGarcke2007, SaibabaKitanidis12}. The inverse of the covariance matrix was approximated in \cite{Ambikasaran16, ambikasaran2013large, bebendorf2003existence}. The $\H$-matrix technique for the parameter estimation was
proposed in \cite{ambikasaran2013large, Ambikasaran16}.  There are many implementations of $\H$-matrices exist: HLIB (\url{http://www.hlib.org/}), $\H^2$ (\url{https://github.com/H2Lib}), HLIBPro (\url{https://www.hlibpro.com/}), and some others. 
In this work, we are using the HLIBPro library. For extended details, we refer to our earlier works \cite{LitvHLIBPro20,LitvHcov19}. 
The data, which we used in this work were generated in the ExaGeoStat library \cite{Sameh18}\\ (https://github.com/ecrc/exageostat) without using $\H$-matrices.
\\
\textbf{Mat\'{e}rn covariance functions:}
We consider the Mat\'{e}rn family \cite{Matern1986a}, which has gained widespread interest in recent years \cite{Guttorp2006a}. The Mat\'{e}rn covariance depends only on the distance $\bh:=\Vert \bs-\bs'\Vert $, where $\bs$ and $\bs'$ are any two spatial locations:
\begin{equation}
\label{eq:MaternCov}
C(\bh;\btheta)=\frac{\sigma^2}{2^{\nu-1}\Gamma(\nu)}\left(\frac{h}{\ell}\right)^\nu K_\nu\left(\frac{h}{\ell}\right)+\tau^2 \bI,
\end{equation}
with parameters $\btheta=(\sigma,\ell,\nu, \tau)^\top$. Here $\sigma^2$ is the variance,
$\tau^2$ the nugget,
$\nu>0$ controls the smoothness of the random field, with larger values of $\nu$ corresponding to smoother fields, and $\ell>0$ the spatial range parameter that measures how quickly the correlation of the random 
field decays with distance. A larger $\ell$ corresponds to a faster decay. ${\cal K}_\nu$ denotes the modified Bessel function of the second kind of order $\nu$.\\
\textbf{Prediction:} Estimating the unknown parameters $\thetab$ is only an intermediate step. Once it is done, the estimation $\widehat{\thetab}\approx \thetab$ is used for prediction at new locations. Let $I_1:=(\bs_{1},\ldots,\bs_{n})$ be locations with known values $\bZ_1$, and $I_2=(\bs_{n+1},\ldots,\bs_{n+m})$ be the new locations with unknown values $\bZ_2=\{Z(\bs_{n+1}),\ldots,Z(\bs_{n+m})\}^\top$ to be predicted.
Here\\
$(Z(\bs_1),\ldots, Z(\bs_{n}),Z(\bs_{n+1},\ldots,Z(\bs_{n+m}))^\top$ is a Gaussian random field indexed by spatial
locations with indices from the index set $(I_1,I_2)$. We assume that vector $(\bZ_1,\bZ_2)$ is zero mean and has a stationary parametric covariance function. After discretisation we can get 
the following block covariance matrix 
\begin{equation}
\label{eq:newcov}
\begin{bmatrix}
\bC_{11} & \bC_{12} \\
\bC_{21} & \bC_{22}
\end{bmatrix},
\end{equation}
where $\bC_{11} \in \mathbb{R}^{n_1 \times n_1}$, $\bC_{12} \in \mathbb{R}^{n_1 \times n_2}$,  $\bC_{21} \in \mathbb{R}^{n_2 \times n_1}$, and  $\bC_{22} \in \mathbb{R}^{n_2 \times n_2}$. Now, the unknown vector $\bZ_2$ can be computed by the following formula \cite{CressieWikle11}
\begin{equation}
\label{eq:prediction}
    \bZ_2=\bC_{21}\bC_{11}^{-1}\bZ_1.
\end{equation}
We can also say that $\bZ_2$ has the conditional distribution with the mean value $\bC_{21}\bC_{11}^{-1}\bZ_1$ and the covariance matrix $\bC_{22} - \bC_{21}\bC_{11}^{-1}\bC_{12}$.

\section{$\H$-matrix approximation of covariance matrices and the log-likelihood}
\label{sec:Hcov}

The $\H$-matrix technique is defined as a recursive partitioning of a given matrix
into sub-blocks. The majority of these sub-blocks are approximated by low-rank matrices on the fly (without computing any dense sub-matrices). And only a minor number of sub-block are calculated as dense matrices without any approximation.
Details about block partitioning and heuristic algorithms used for low-rank approximation are not so trivial. Therefore, we skip them here and refer to \cite{khoromskij2009application,LitvHcov19,LitvHLIBPro20}.

The $\H$-matrix approximation error depends on the type of the covariance matrix, its smoothness, covariance length, computational geometry, nugget, and the dimensionality of the problem. For some matrices, the problem may become ill-posed since even tiny perturbations in the covariance matrix $\bC(\btheta)$ may result in considerable perturbations in the log-determinant and the log-likelihood. The usage of $\tau^2\bI$ regularisation helps partially to resolve this issue. 

\textbf{Storage and complexity.}
We let $\bC(\btheta)\in \mathbb{R}^{n \times n}$ be approximated by an $\H$-matrix $\widetilde{\bC}(\btheta;k)$ or $\widetilde{\bC}(\btheta;\varepsilon)$. In the first case we fix the maximal rank $k$ in each sub-block (the approximation accuracy will vary from sub-block to sub-block). In the second case we fix the accuracy $\varepsilon$ in each sub-block (the ranks of sub-blocks will vary). The $\H$-Cholesky decomposition of $\widetilde{\bC}(\btheta;k)$ costs $\mathcal{O}(k^2 n\log^2 n)$. The solution of
the linear system $\widetilde{\bL}(\thetab;k)\bv(\btheta)=\bZ$ costs $\mathcal{O}(k^2 n\log^2 n)$.
The log-determinant $\log \mydet{\widetilde \bC(\btheta;k)}=2\sum_{i=1}^n \log \{\widetilde L_{ii}(\btheta;k)\}$ is available for free. 
The cost of computing the log-likelihood function $\widetilde{\LL}(\thetab;k)$ is $\mathcal{O}(k^2 n\log^2 n)$ and the cost of computing the MLE $\widehat{\btheta}$ in $m$ iterations is ${\mathcal{O}(m k^2 n\log^2 n)}$.\\
\textbf{Maximization of the log-likelihood.}
To maximize $\widetilde{\LL}(\thetab;k)\approx \LL(\thetab)$ we use the Brent-Dekker method \cite{Brent73, BrentMethod07}. It is implemented in the GNU Scientific library \url{https://www.gnu.org/software/gsl/}. The  Brent-Dekker algorithm first uses the fast-converging secant method or inverse quadratic interpolation to maximize $\widetilde{\LL}(\thetab;\cdot)$. If those do not work, then it returns to the more robust bisection method. In the following we will call this optimization procedure $\H$-MLE. It iteratively computes parameter $\thetab$ where the maximum of $\widetilde{\LL}(\thetab;\cdot)$ is achieved.

Additionally to the $\H$-Cholesky factorisation $\bC(\btheta)=\bL(\btheta)\bL(\btheta)^\top$, we implemented a more stable factorisation $\bL(\btheta) \bD(\btheta) \bL^\top(\btheta)$, which avoids extracting square roots of diagonal elements. Both factorizations are connected via $\bL \bD \bL^\top=(\bL\bD^{1/2})(\bL\bD^{1/2})^\top$. Very small negative diagonal elements can appear due to, e.g., the rounding off error.

The computation of $\widehat{\btheta}$ depends on the number of iterations in the optimization algorithm and the used threshold ($10^{-4}$ in our experiments). The maximal number of iterations we used was 400. We may need more depending on the initial guess and the threshold. The running times are listed in Table~\ref{t:HMLE-times}.\\
\textbf{$\H$-matrix approximation error analysis.} For multiple numerical tests we refer to our earlier works \cite{LitvHcov19,LitvHLIBPro20,khoromskij2009application}. There the reader can find numerical errors for $\bC$, the Cholesky factor $\bL$, and the log-likelihood $\LL$. The $\H$-matrix approximation accuracy of the Cholesky factor and the inverse depends on the condition number of $\bC$. The prediction accuracy can be estimated as follows
\begin{align*}
    \Vert \bZ_2 - \widetilde{\bZ}_2 \Vert 
    &= \Vert \bC_{21}\bC_{11}^{-1}\bZ_1 - \widetilde{\bC}_{21}\widetilde{\bC}_{11}^{-1}\bZ_1 \Vert\\
    &= \Vert \bC_{21}\bC_{11}^{-1}\bZ_1 - \widetilde{\bC}_{21}\bC_{11}^{-1}\bZ_1 +  \widetilde{\bC}_{21}\bC_{11}^{-1}\bZ_1 - \widetilde{\bC}_{21}\widetilde{\bC}_{11}^{-1}\bZ_1 \Vert\\
    &\leq \Vert \bC_{21}\bC_{11}^{-1}\bZ_1 - \widetilde{\bC}_{21}\bC_{11}^{-1}\bZ_1 \Vert + \Vert \widetilde{\bC}_{21}\bC_{11}^{-1}\bZ_1 - \widetilde{\bC}_{21}\widetilde{\bC}_{11}^{-1}\bZ_1 \Vert\\
    &\leq \Vert \bC_{21} - \widetilde{\bC}_{21} \Vert \cdot \Vert \bC_{11}^{-1} \Vert \cdot \Vert \bZ_1 \Vert + \Vert \widetilde{\bC}_{21} \Vert \cdot \Vert \bC_{11}^{-1} - \widetilde{\bC}_{11}^{-1} \Vert \cdot \Vert \bZ_1 \Vert
   \end{align*}
Now we see that the quality of the prediction depends on the quality of the $\H$-matrix approximation of  matrices $\bC_{21}$ and $\bC^{-1}_{11}$, i.e.
the norms $\Vert \bC_{21} - \widetilde{\bC}_{21} \Vert $ and $\Vert \bC_{11}^{-1} - \widetilde{\bC}_{11}^{-1} \Vert$. In our earlier works \cite{LitvHcov19,LitvHLIBPro20,khoromskij2009application}, we demonstrated the error decay for $\bC$ and $\bC^{-1}$.
%
%
%
\section{Prediction Errors}
We used the Mean Loss Efficiency (MLOE), the Mean Misspecification of the Mean Square Error (MMOM), and the Root Mean Square Error (RMSE) as in the 2021 KAUST Competition on Spatial Statistics for Large Datasets \cite{KAUSTcomp}:
\begin{equation}
\label{eq:MLOE}
\mbox{MLOE}:=\frac{1}{M}\sum_{j=1}^{M} \left( \frac{\EXPT{(\hat{Z}_a(\bs_j) - Z(\bs_j))^2} }{\EXPT{(\hat{Z}_t(\bs_j) - Z(\bs_j))^2}} - 1 \right),
\end{equation}
\begin{equation}
\label{eq:MMOM}
\mbox{MMOM}:=\frac{1}{M}\sum_{j=1}^{M} \left( \frac{\EXPA{(\hat{Z}_a(\bs_j) - Z(\bs_j))^2} }{\EXPT{(\hat{Z}_a(\bs_j) - Z(\bs_j))^2}} - 1 \right). 
\end{equation}
Here $(\bs_1,\ldots,\bs_M):=\mathcal{J}$ is a fixed subset of $M<n$ randomly-chosen locations. For numerical purposes $M$ was chosen to be equal $1000$. $\hat{Z}_t(\bs_j)$ and $\hat{Z}_a(\bs_j)$ are respectively kriging prediction at $\bs_j$ using the true and approximated
model (plugging in the true parameters and estimated parameters in the covariance
function), and $\EXPT{\cdot}$ and $\EXPA{\cdot}$ are respectively the expectation using the true and approximated model. We refer to \cite{KAUSTcomp} for more details.

MLOE gives us an understanding of the average loss of prediction efficiency when the approximated model is used to predict
instead of the true model. MMOM presents the average misspecification of the mean
square error when calculated under the approximated model.

The RMSE error was used to evaluate the prediction accuracy
\begin{equation}
\label{eq:RMSE}
\mbox{RMSE} = \sqrt{\frac{1}{n_t} \sum_{i=1}^{n_t} \left( \hat{Z}(\bs_i) - Z(\bs_i)\right)^2},    
\end{equation}
where $\hat{Z}(\bs_i)$ and $Z(\bs_i)$ are respectively the predicted and true realization values at the
location $\bs_i$ in the testing dataset, and $n_t$ is the total number of locations in
the testing dataset.
\section{Machine learning methods to make predictions}
\begin{figure}[h!]
\centering
\includegraphics[width=0.7\textwidth]{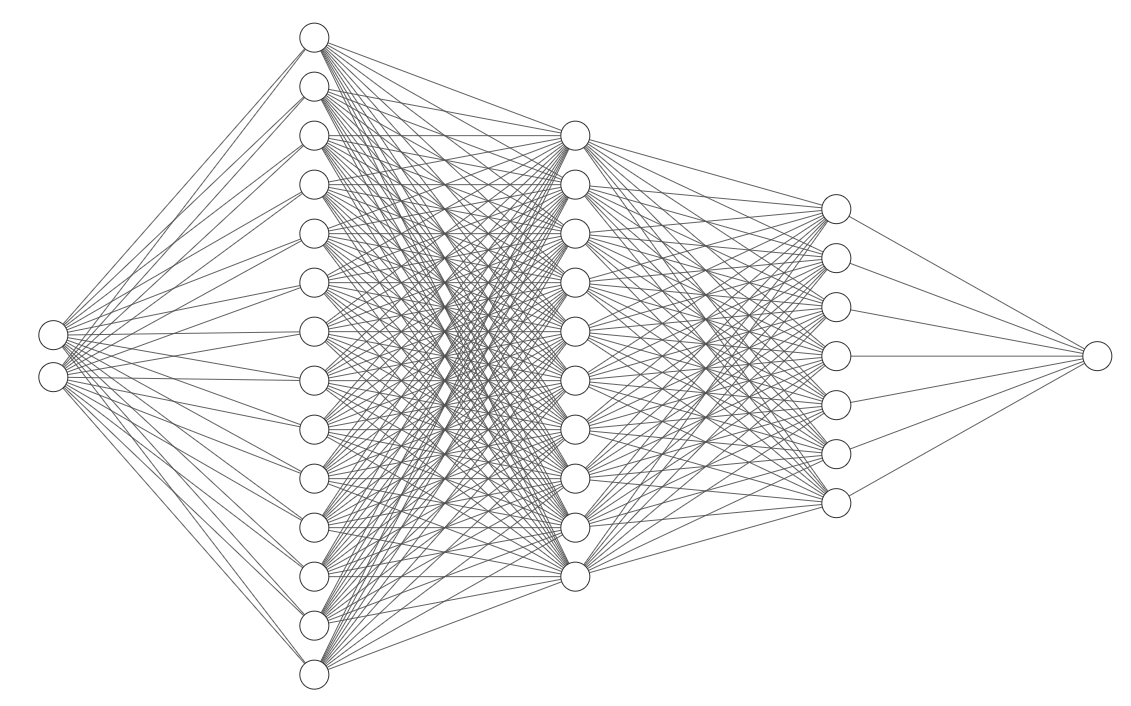}
\caption{Example of FCNN architecture. The input layer consists of two neurons (input feature dimensionality), and the output layer consists of one neuron (predicted feature dimensionality). Three hidden layers with different numbers of neurons are presented.}
\label{fig:FCNN}
\end{figure}

Machine learning is aimed at building a model of data automatically from the observations. The obtained predictions can be considered as a baseline for comparison with other forecasting methods in which some additional information on the studied process is used. In the following, we tried three methods:\\
\textbf{k-nearest neighbours (kNN):} This method belongs to classical non-parametric family of statistical machine learning methods and follows a simple idea: for each data point $x$ for which one needs to  predict its output $\hat y$, find its $k$ nearest neighbors $x_{1},\dots,x_k$ with respect to some metrics, and set $\hat y=\frac{1}{k} \sum_{i=1}^k y_i $, where $y_i$ is the observed value of the response for point $x_i$.  The value of $k$ should be determined in the best way, for example, by cross-validation procedure or using an independent test sample for error estimation.\\
\textbf{Random Forest (RF):} Random Forest (RF) is another popular machine learning method in which a large number of decision (or regression) trees are generated independently on random sub-samples of data. The final decision for $x$ is calculated over the ensemble of trees by averaging the predicted outcomes. The method is theoretically well-substantiated and gives state-of-the-art results in many practical tasks, especially in the presence of many irrelevant features describing the observed data.   \\
\textbf{Deep Neural Network (DNN):} Methods of this broad class are based on the artificial neural network paradigm, which models the functioning of neurons in the brain. In our study, we use a fully connected neural network (FCNN) which includes several fully connected layers, i.e., connecting each neuron in a layer to every neuron in the next layer (see an example in Figure~\ref{fig:FCNN}). Mathematically speaking, the input feature vector transformation performed with each layer can be presented as a matrix-vector multiplication, where the matrix elements are neuron connection weights. Each layer is followed by a non-linear activation unit. FCNN training procedure consists of finding neurons' connection weights for which the quality metric takes the best value (usually by gradient descent technique).\\
Each ML method needs a fine-tuning stage to optimize its hyperparameters or architecture. The best value of $k$ and the distance metric should be determined for kNN. For RF, one needs to set the number of trees in the ensemble, tree complexity, and splitting criterion. For FCNN, one needs to optimize the number of layers and neurons in each layer. For some other DNN ( kNN and RF) parameters, we use default Matlab, Scikit-learn, or Tensorflow settings. For other hyperparameters (such as $k$ for kNN, the number of trees for RF, or the number of layers and hidden units for FCNN), we minimized the root mean squared error (RMSE) metric using a validation sample repeatedly obtained by random sub-sampling of data in the proportion 1:9. We used a candidate set of $k$ values in the interval $\{1,\dots,20\}$, number of trees in the interval $\{100,\dots,150\}$ and examined some variants of FCNN architecture for different number of hidden layers in the interval [3, 10] (having 50-100 neurons in each hidden layer).
\section{Numerical results, obtained by the $\H$-MLE method}
\textbf{Datasets:} For all tests below we used datasets from the statistical competition \cite{KAUSTcomp}.
The spatial domain is the unit square $\mathcal{D}:=[0.0, 1.0] \times [0.0, 1.0]$. Each dataset includes $90\%$ of training samples and $10\%$ testing samples, where prediction should be made. 

Below we will describe the following numerical tests: 1a, 1b, 2a, and 2b.

Tests 1a, 1b, and 2a contain $100,000$ locations, and Test 2b contains $1,000,000$ locations. The training datasets and datasets for predictions are taken from \cite{KAUSTcomp}.\\
\textbf{Hardware:} 
For the $\H$-MLE method we used a parallel cluster with two Intel Xeon Gold 6144 processors. Each processor has 8 cores (16 threads) with 3.5GHz and 384GB RAM in total.\\
\textbf{Software:} All $\H$-MLE numerical results are reproducible. We invite the reader to install HLIBPro-2.9 (from \url{www.hlibpro.com}), download our code from \url{https://github.com/litvinen/large_random_fields.git} and play with it.\\

Parameters identified in Test 1a are used for the prediction in Test 1b. 
Parameters identified in Tests 2a and 2b are used for prediction in Tests 2a and 2b, respectively.

After we identified all parameters and did all predictions, we uploaded all these data to the competition webpage\cite{KAUSTcomp} and the organisers of that competition computed for us the approximation errors. These errors are listed in Tables~\ref{t:S1a_params} and \ref{t:S2_RMSE_errors_kNN}.

\subsection{Tests 1a and 1b: Parameter identification and prediction, 8 datasets with $n=90,000+10,000$}

In the Test 1a, there are 16 given datasets from different zero-mean stationary isotropic Gaussian
random fields with a Mat\'ern covariance.
The training dataset consists of $90,000$ randomly distributed locations and associated observations at these locations. The task is to infer four unknown parameters of the Mat\'ern covariance function
shown in \eqref{eq:MaternCov} for each dataset.

To avoid negative intermediate values for these parameters, in the following we assume that:
\begin{displaymath}
\label{eq:log}
      \sigma=\frac{2.0}{1.1^{\sigma_0}},\quad
      \ell = \frac{1.0}{1.5^{\ell_0}},\quad
\nu = \frac{1.0}{1.2^{\nu_0}}, \quad
\tau = \frac{1.0}{2.0^{\tau_0}},
\end{displaymath}
where $\sigma_0$, $\ell_0$, $\nu_0$, $\tau_0$ the new parameters to be identified
by the optimization algorithm. As the initial guess, we took $(\sigma_0, \ell_0, \nu_0, \tau_0)=(2,2,1,15)$. If we saw that we were wrong with these values (too many iterations were needed), we rerun the optimization algorithm with some new values.
The advantage of this ``log"-representation is that the auxiliary values $\sigma_0$, $\ell_0$, $\nu_0$, $\tau_0$ are allowed to take negative values, whereas $\sigma$, $\ell$, $\nu$, $\tau$ not. Negative values may appear during iterations in the MLE optimization procedure.

Table~\ref{t:S1a_params} contains 8 solutions for 8 given datasets \cite{KAUSTcomp}. The 1st column contains the dataset index, columns 2,3,4 and 5 contain values of $(\sigma^2,\ell,\nu,\tau^2)$ respectively. The column 6,7,8, and 9 contain the true values $(\hat{\sigma}^2,\hat{\ell},\hat{\nu},\hat{\tau}^2)$ respectively. The columns 10 and 11 contain the MLOE and MMOM errors. The 12th columns contains the RMSE error (as defined in Eq.~\ref{eq:RMSE}).
One can see that in some rows the estimated parameter values are very close to the true values, but in some not.
The reason is that the derivative of the log-likelihood function at the point
$(\sigma,\ell,\nu,\tau)$ is almost zero (is equal to our threshold $10^{-4}$), and our optimization algorithm indicates this point as the maximum. To improve the estimate, we should iterate longer. Later, in Test 1b, the estimated parameters are used for the prediction, and one can see that our predictions are reasonable.

\begin{table}[h!]
\centering
\begin{footnotesize}
\begin{tabular}{|c|c|c|c|c||c |c|c|c||c|c|c|} \hline
 dataset & $\sigma^2$ & $\ell$   & $\nu$    & $\tau^2$  
 & $\hat{\sigma}^2$ & $\hat{\ell}$   & $\hat{\nu}$    & $\hat{\tau}^2$ & MLOE & MMOM & RMSE \\ \hline
1 & 0.29 & 0.0106 & 2.471 & 2.5e-14   &1.5& 0.0175& 2.3& 0 &2.2e-2 & 4.8e-1 & 4e-3\\ \hline
2 & 1.762 & 0.0223 & 1.501 & 1.1e-14   &1.5&0.0211&1.5& 0 & 1.8e-4 & 8.8e-2 & 2.4e-2\\ \hline    
3 & 1.478& 0.0305, & 0.600 &  1.0e-10   &1.5&0.031&0.6& 0 & 2.0e-6 & 8.0e-3 & 0.23\\ \hline
{4} & 1.09 &0.0176,  &  1.522 &  7.0e-14    &1.5&0.0526&2.3&0 & 1.8 & 3566 & 5.6e-4\\ \hline
5 & 0.95& 0.0781       &0.714 &  1.3e-13    &1.5&0.0632& 1.5 & 0 & 2.2e-1 & 100 & 5.4e-3    \\ \hline
6 & 1.32 & 0.0826 & 0.601 & 1.22e-27   &1.5&0.0928&0.6&0    & 5.4e-4 & 5.3e-2 & 0.12\\ \hline
7 & 2.38& 0.4370 & 0.795 &  2.47e-8    &1.5&0.1686&1.5& 0     & 2.5e-1 & 164 & 2e-3\\ \hline
8 & 1.2& 0.2043 & 0.601 & 4.1e-17  &1.5&0.2475&0.6&0     & 2.8e-3 & 6.4e-2 & 6.6e-2\\ \hline
\end{tabular}
\end{footnotesize}
\caption{Results of the $\H$-MLE method. Comparison of the identified (columns 2-5) and true parameters (columns 6-9).}
\label{t:S1a_params}
\end{table}
%
One can see that for some datasets (e.g., 4,5,7), the MLOE and MMOM errors are large. We would not say that the $\H$-matrix method failed since the optimization algorithm's accuracy to compute the MLE estimate was $10^{-4}$. We think that the problem is ill-posed and contains multiple solutions, i.e., there are many points $\thetab$ where the derivative of $\LL$ is almost zero. Here ``almost" means smaller than $10^{-4}$.

\begin{figure}[htbp!]
\centering
\includegraphics[width=0.45\textwidth]{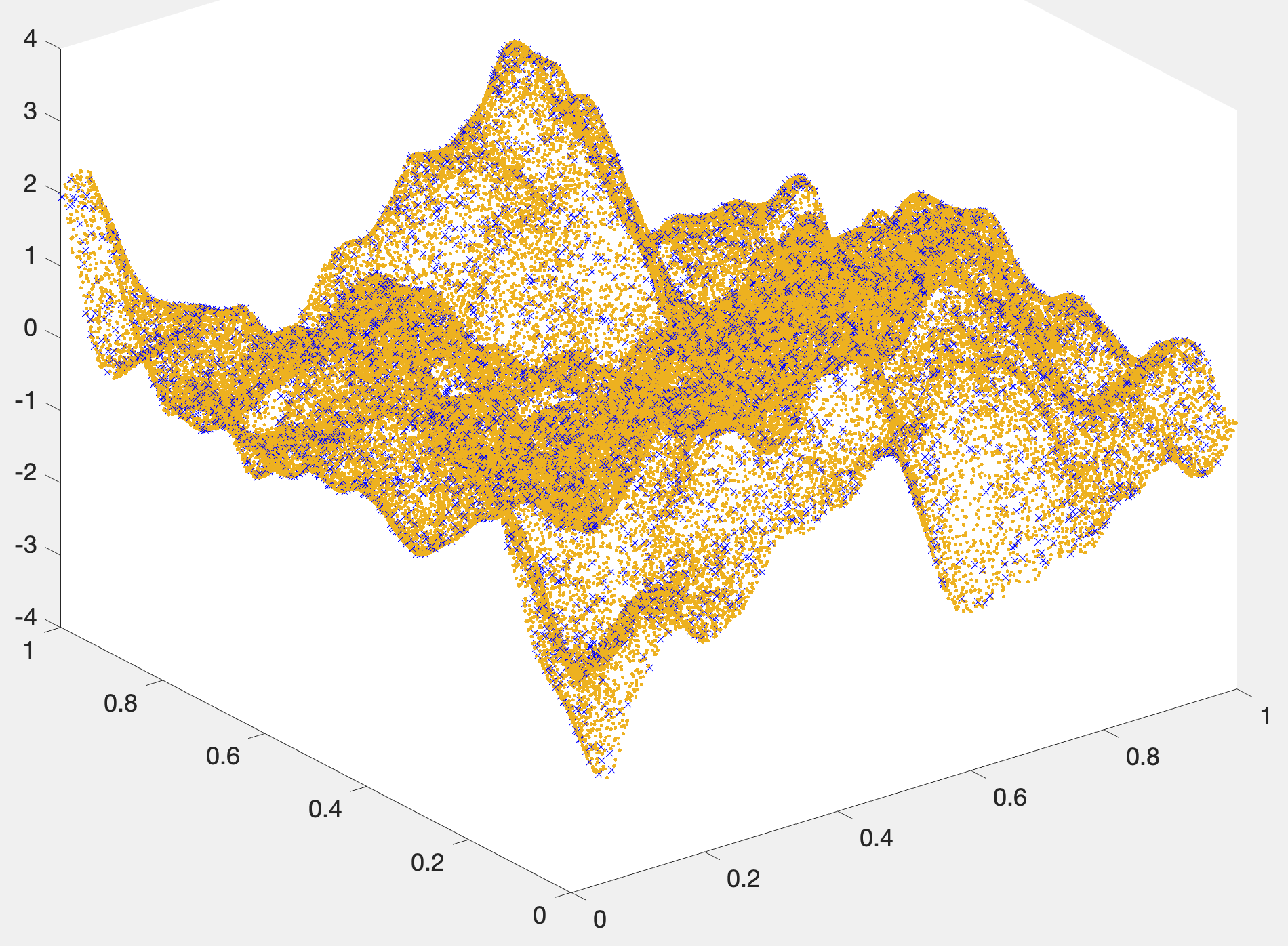}
\includegraphics[width=0.45\textwidth]{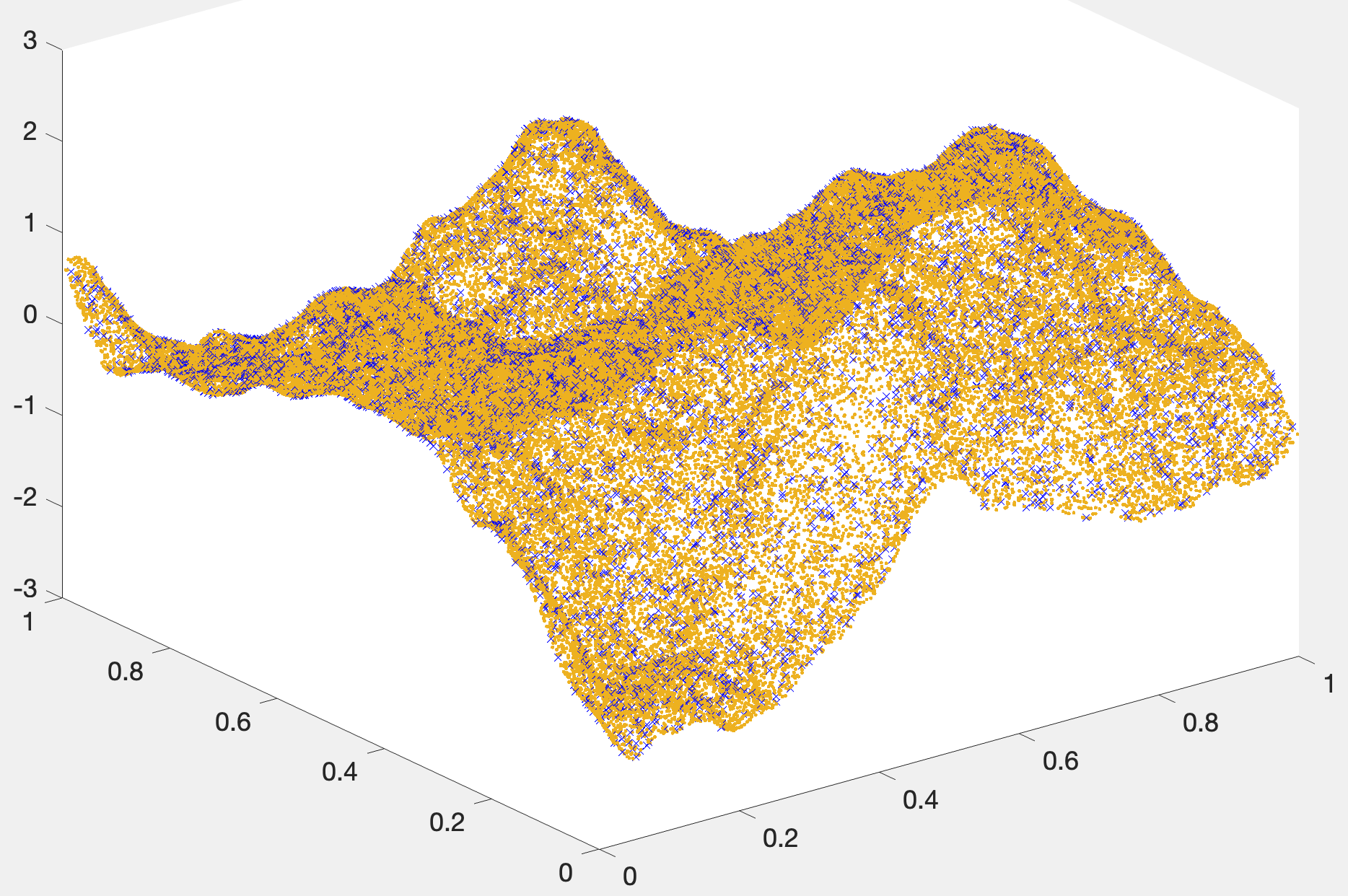}
\caption{Test 1b, datasets 4 and 7: Prediction obtained by the $\H$-MLE method. The yellow points at $90,000$ locations were used for training and the blue points were predicted in $10,000$ locations. Although the identified parameters for these datasets were far away from the true values (see rows 4 and 7 in Table~\ref{t:S1a_params}), one can still observe a very good alignment of yellow and blue points.}
\label{fig:S1b_difficult_dataset_4_7}
\end{figure}

Equation~\ref{eq:prediction} was used to do prediction at the new $10,000$ locations. Figure~\ref{fig:S1b_difficult_dataset_4_7} (left and right) visualize datasets 4 and 7 (see the 4th and 7th rows in Table~\ref{t:S1a_params}), where the estimated parameters are far away from the true values (see also the last three columns in Table~\ref{t:S1a_params}).  As we can see $90,000$ given measurements (yellow points) and $10,000$ predicted (blue points) for both datasets are very good aligned.

\subsection{Test 2a: Parameter identification and prediction, $n=90,000+10,000$}
There are two datasets in this experiment. Each dataset contains $90,000$ measurements to identify unknown parameters (training of the statistical model). In later experiments these unknown parameters will be used to predict unknown values in new $10,000$ locations.
The identified parameters for both datasets are listed in Table~\ref{tab:test2a}. The columns 2,3,4,5 contain the values $(\sigma^2,\ell,\nu,\tau^2)$ respectively, the 6th column the value $\LL(\sigma^2,\ell,\nu,\tau^2)$, and columns 7,8,9,10 contain the true values $(\hat{\sigma}^2,\hat{\ell},\hat{\nu},\hat{\tau}^2)$ respectively. These true values were obtained from organisators after the competition \cite{KAUSTcomp} finished.

Parameters $\nu$ and $\tau$ were identified well, but $\sigma^2$ and $\ell$ not. We see two possible reasons for this. The first reason is that the true model was not Gaussian. The second reason is the insufficient threshold $10^{-4}$ in the MLE optimization algorithm.
Initially, the organizers did not provide any additional information about the utilized model. Here, we actually tested how the Gaussian model approximates the Tukey g-and-h random model \cite{Genton17}. Meaning, that both datasets in Task-2a were univariate non-Gaussian spatial datasets, which were generated by the Tukey $g$-and-$h$ random fields. These fields generalize Gaussian random fields $Z(\bs)$. The Tukey $g$-and-$h$ random process $T(\bs)$ is defined by marginal transformation at each location $\bs$ as follows:
\begin{equation}
    T(\bs):= \xi + \omega\cdot\frac{\exp(g\cdot Z(\bs))-1}{g}\cdot \exp\left(\frac{hZ^2(\bs)}{2}\right ),
\end{equation}
where $\xi$ and $\omega$ are the location and scale parameters. The parameter $g$ defines the skewness and $h\geq 0$ the tail-heaviness. Two different pairs of $(g,h)$ were chosen, which simulate medium and strong deviation from Gaussian random fields. These parameters $\xi$, $\omega$, $g$, $h$ used for generating both datasets are listed in Table~\ref{tab:test2a}. 

\begin{table}[h!]
\centering
\begin{tabular}{|c|c|c|c|c|c|||c|c|c|c|c|c|c|c|}\hline
 data  &$\sigma^2$ & $\ell$      & $\nu$    & $\tau^2$       & $\mathcal{L}$ &$\hat{\sigma}^2$ & $\hat{\ell}$      & $\hat{\nu}$    & $\hat{\tau}^2$ & $\xi$ & $\omega$ & $g$ & $h$ \\ \hline
1 & 7.7 & $0.07$ & $1.037$ & $1.6e-15$ & $9.6e+4$ & 1 & 0.1 & 1 & 0 &1&2&0.2&0.2\\ \hline
2 & 31 & $0.047$ & $1.066$ & $4.0e-14$ & $0.5e+4$ & 1 & 0.1 & 1 & 0 &1&2&0.5& 0.3\\ \hline
\end{tabular}
\caption{Results of the $\H$-MLE method. Comparison of the obtained parameter values with the true values for Test 2a, $n=90,000$.}
\label{tab:test2a}
\end{table}

\begin{figure}[h!]
\centering
\includegraphics[width=0.45\textwidth]{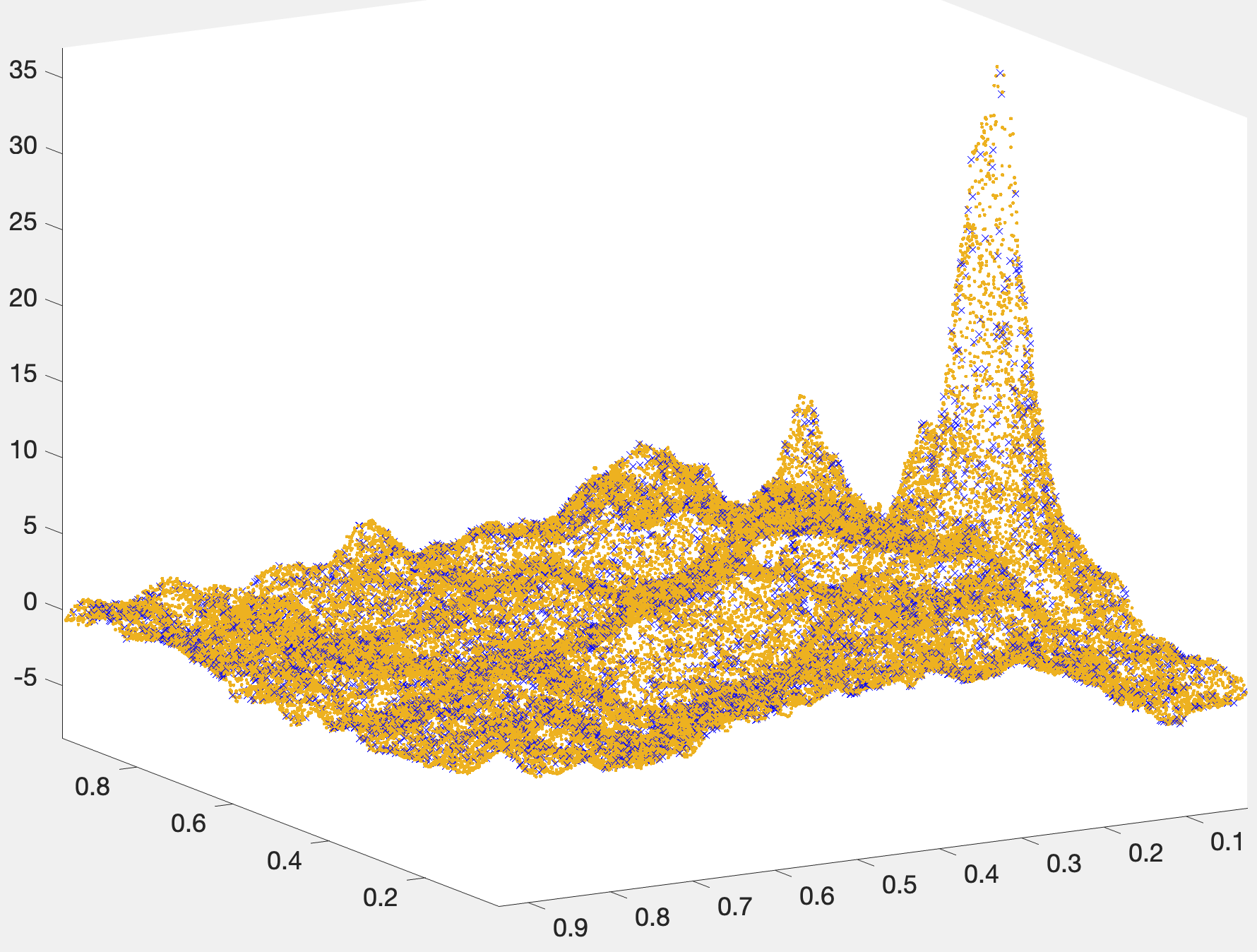}
\includegraphics[width=0.45\textwidth]{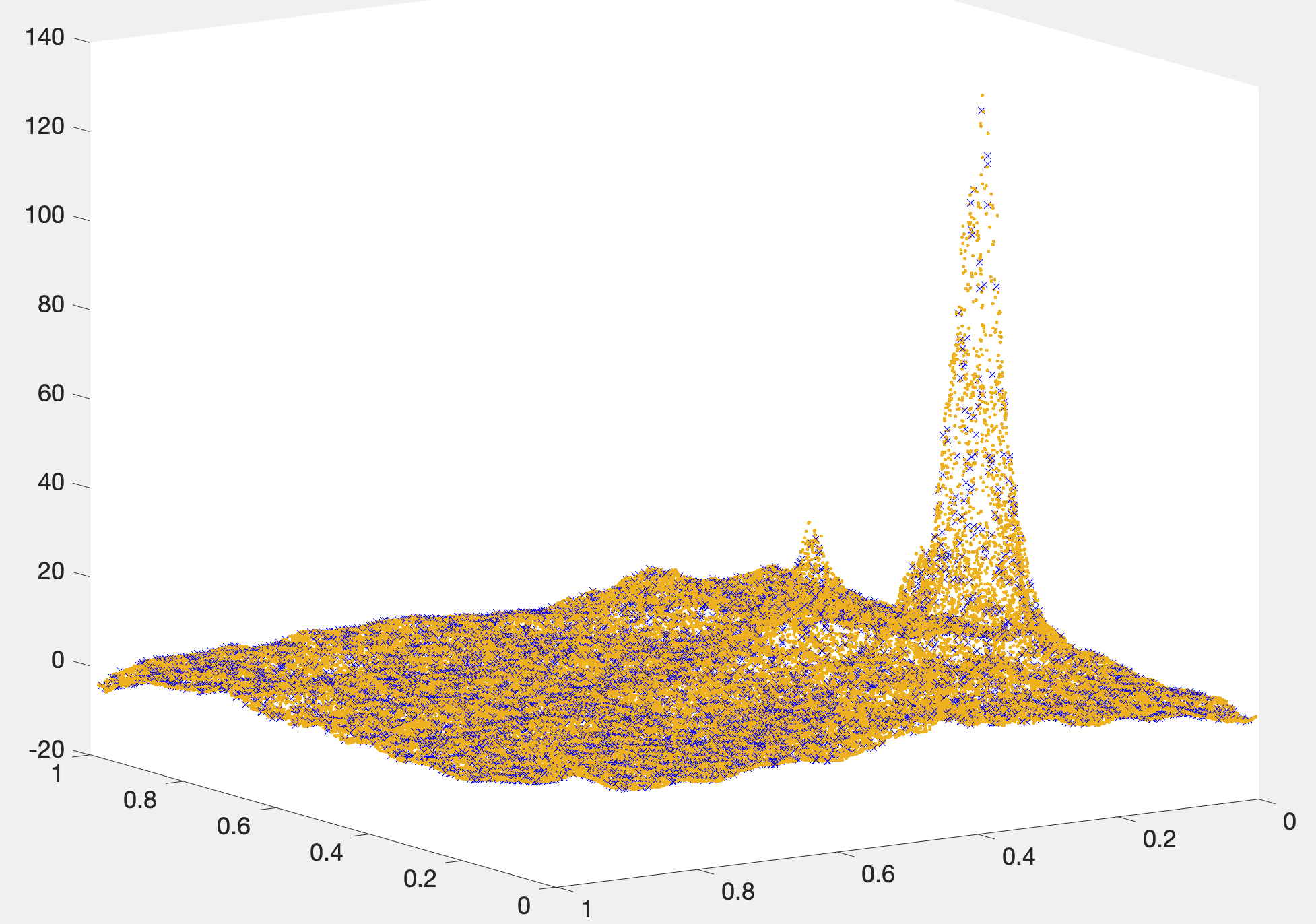}
\caption{Test 2a, datasets 1 and 2: Prediction obtained by the $\H$-MLE method. The yellow points at $90,000$ locations were used for training and the blue points were predicted in $10,000$ locations. One can see a very good alignment of yellow and blue points on both figures.}
\label{fig:S2a_dataset1}
\end{figure}

Figures~\ref{fig:S2a_dataset1} (left and right) show predictions obtained by the $\H$-MLE method. The yellow points at $90,000$ locations were used for training and the blue points were predicted in $10,000$ new locations. One can see a very good alignment of yellow and blue points on both pictures.

%
%
%
%
%
%
\subsection{Test 2b: Parameter identification and prediction, $n=900,000+100,000$}
There are two datasets in this experiment. Each dataset contains $900,000$ measurements to identify unknown parameters (training of the statistical model). Later this model is used to predict unknown values in new $100,000$ locations.
The identified parameters are listed in Table~\ref{tab:test2b-large} for two datasets. The columns 2,3,4,5 contain the values $(\sigma^2,\ell,\nu,\tau^2)$ respectively, the 6th column the value $\LL(\sigma^2,\ell,\nu,\tau^2)$, and columns 7,8,9,10 contain the true values $(\hat{\sigma}^2,\hat{\ell},\hat{\nu},\hat{\tau}^2)$ respectively. These true values were obtained from organisators after the competition \cite{KAUSTcomp} finished.

\begin{table}[h!]
\centering
\begin{tabular}{|c|c|c|c|c|c|||c|c|c|c|} \hline
 data &$\sigma^2$ & $\ell$      & $\nu$    & $\tau^2$       & $\mathcal{L}$ &$\hat{\sigma}^2$ & $\hat{\ell}$      & $\hat{\nu}$    & $\hat{\tau}^2$  \\ \hline
1 & 3.72 & 1.143830 & 0.94636 & 4.4e-3 & 3.6e+5 &1.5 & 0.0632 & 1.5 & 0     \\ \hline
2 &  0.92 & 0.012496 & 1.30867 & 8.5e-9 & 1.8e+6 &1 & 0.1 & 1 & 0\\ \hline
\end{tabular}
\caption{$\H$-MLE method. Comparison of the obtained parameter values with the true values for Test 2b, $n=900,000$.}
\label{tab:test2b-large}
\end{table}
Table~\ref{t:HMLE-times} summarizes the computational times for the $\H$-MLE method.
We note that this computing time varies a lot for the parameter identification task because it depends on the initial guess in the optimization algorithm. If the initial guess lies very close to the (unknown) true value of $\thetab$, then only a few iterations are needed. If not, then a few hundred iterations and the computing time may increase by a factor of 10.

\begin{table}[h!]
\centering
\begin{tabular}{|c|c|c|c|c|} \hline
Datasets/ & 1a & 1b  & 2a    & 2b \\ 
Tasks & parameter  & prediction & parameter inference, & parameter inference, \\ 
 & inference &  & prediction    &  prediction \\ \hline
$\H$-MLE &&&& \\ 
comp. time (sec.) &  360-3600& 60 &  180-3600, 120& 3600-36000, 600  \\ \hline  
\end{tabular}
\caption{Computing time for the parameter inference and for the prediction, $\H$-MLE method.}
\label{t:HMLE-times}
\end{table}
%
%
%

\begin{figure}[htbp!]
\centering
\includegraphics[width=0.45\textwidth]{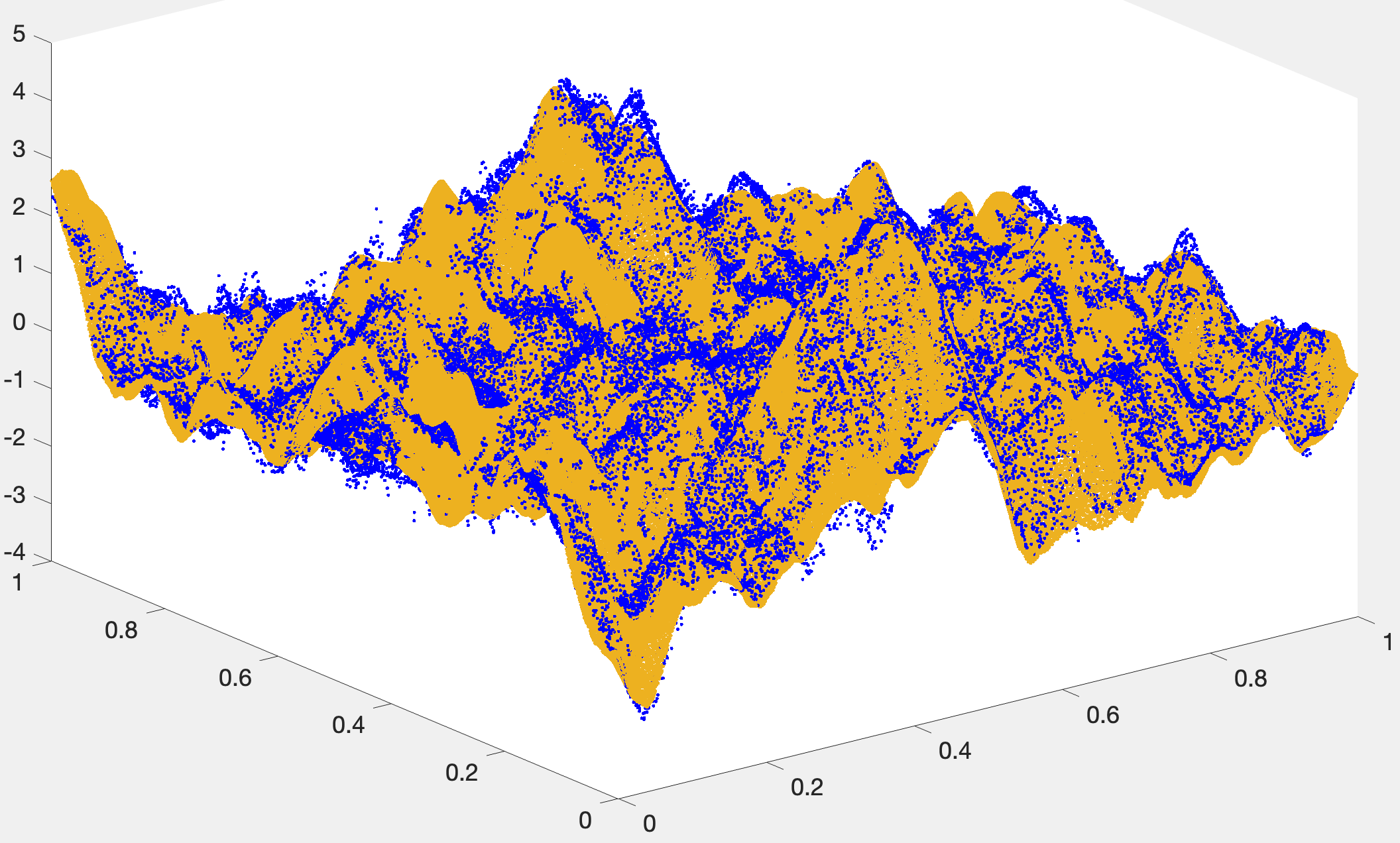}
\includegraphics[width=0.45\textwidth]{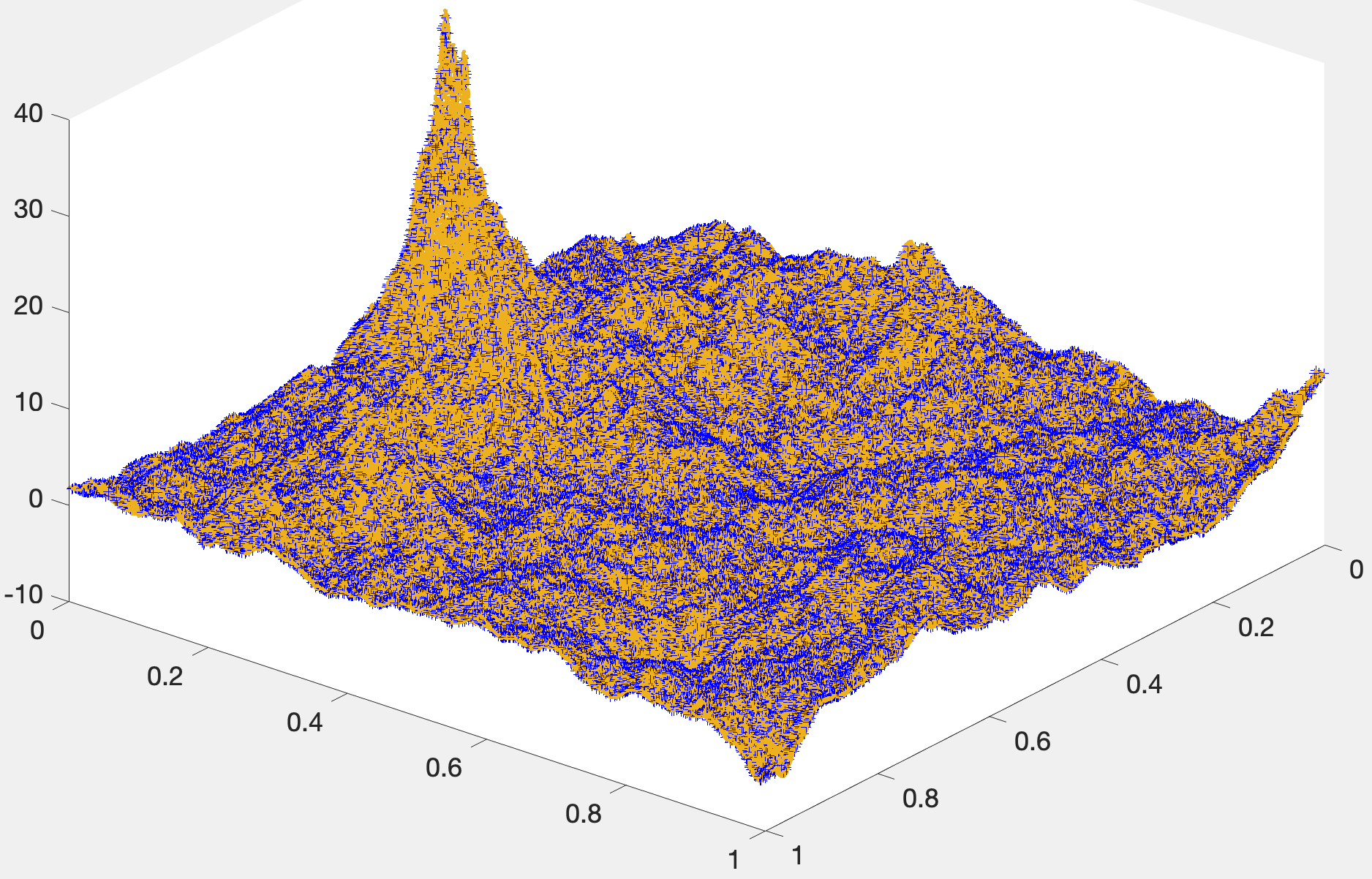}
\caption{Test 2b, datasets 1 and 2: Prediction obtained by the $\H$-MLE method. The yellow points at $900.000$ locations were used for training and the blue points were predicted in $100.000$ locations. One can see a very good alignment of yellow and blue points (on the right) and slightly different values on the left.}
\label{fig:S2b_dataset1}
\end{figure}

%
Figures~\ref{fig:S2b_dataset1} (left and right) show predictions obtained by the $\H$-MLE method. The yellow points at $900.000$ locations were used for training and the blue points were predicted at $100.000$ new locations. One can see a very good alignment of yellow and blue points (on the right) and slightly different values on the left.

\newpage
\section{Numerical results, obtained by the machine learning methods}
\begin{figure}[h!]
\centering
\includegraphics[width=0.49\textwidth]{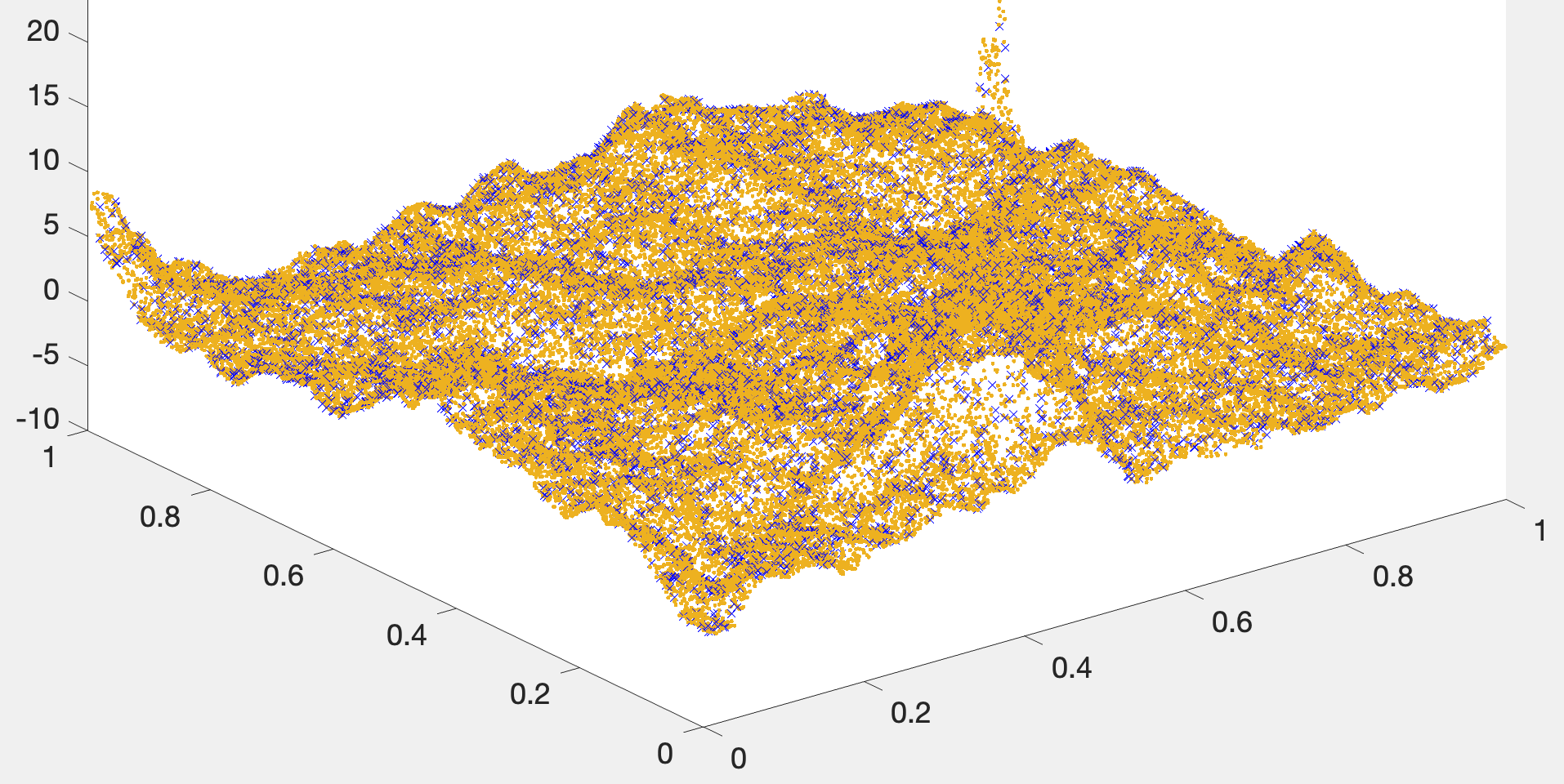}
\includegraphics[width=0.49\textwidth]{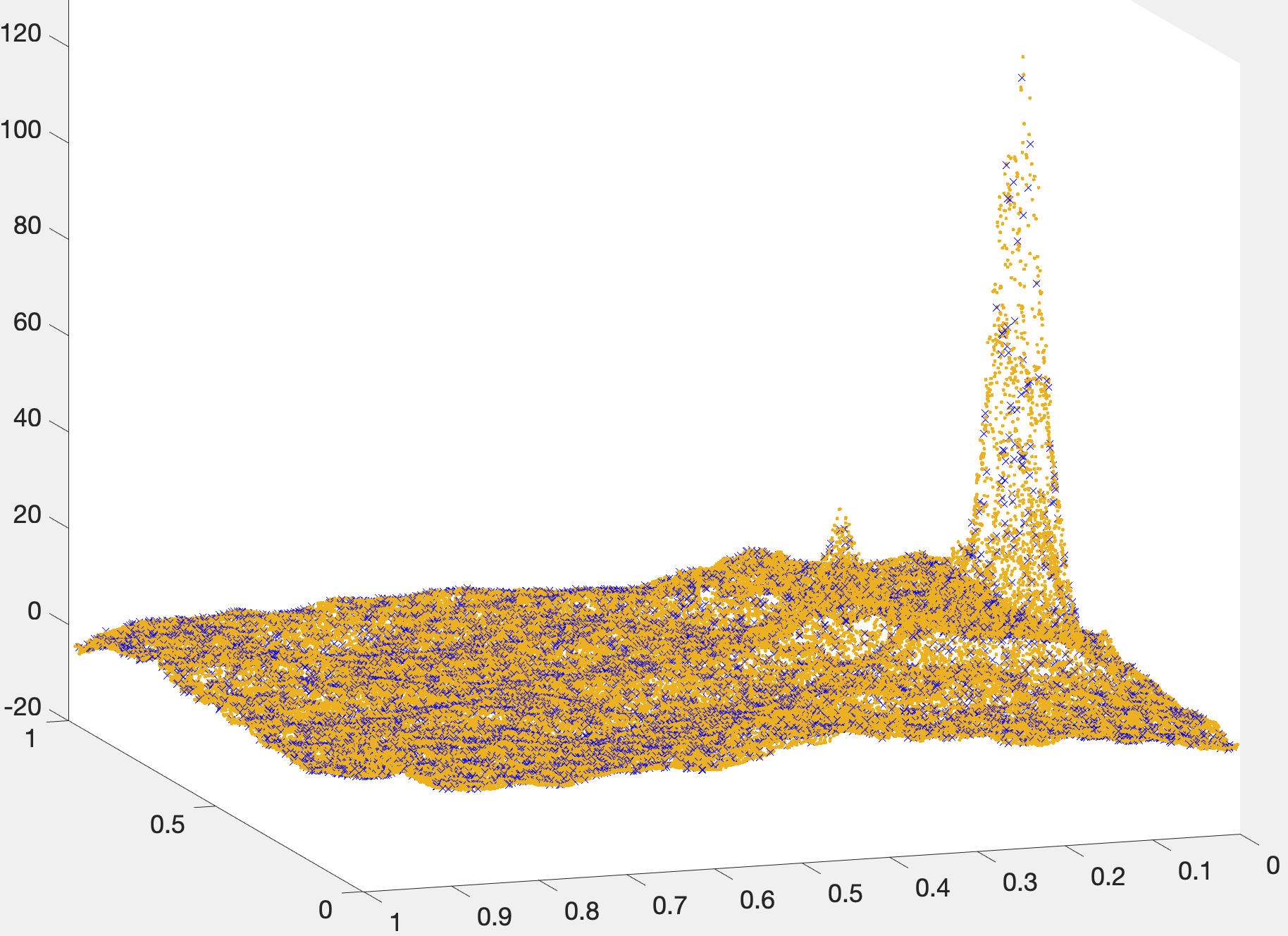}
\caption{Test 2a, datasets 1 and 2: Prediction obtained by the kNN method. The yellow points at $90.000$ locations were used for training and the blue points were predicted at $10.000$ new locations. One can see a very good alignment of both.}
\label{fig:Test2a-1_kNN}
\end{figure}

To run the kNN method we used a usual notebook with Intel i5-9300 CPU, 2.40GHz and 8 GB RAM. We perform Monte-Carlo simulations, in which we repeatedly split the given dataset on training and testing subsamples and average the obtained prediction error estimates over all runs. In our experiments on both datasets in Test 2a, the kNN method has shown the best Monte-Carlo cross-validation results (over 100 runs) in comparison with other used ML methods. Predictions for datasets 1 and 2 from Test 2a are shown in Fig.~\ref{fig:Test2a-1_kNN}. 

Running the kNN method with different $k$, we found out that $k=3$ is optimal for Test 2a, and  $k=7$ for Test 2b.
Trying different numbers of trees in the random forest method, we defined that an ensemble of 120 regression trees is optimal. Further, we have designed the FCNN architecture with 7 hidden layers with 100 neurons in each layer. The number of training epochs is 500, and the batch size equals 10000. We use $\mbox{tanh}(\cdot)$ activation function and Adam optimizer. The average calculation time is $0.07$ sec. for kNN (k-d tree was used to speed up calculations), $12$ sec. for RF and $173$ sec. for FCNN. Because of its efficiency, we decided to run only the kNN method for the prediction in Test 2b.

Table~\ref{t:S2_RMSE_errors_kNN} contains the RMSE errors for all methods (defined in Eq.~\ref{eq:RMSE}). Note that the dataset2 from Test 2b is sampled from the same random field as the dataset1 from Test 2a. The dataset1 from Test 2b is sampled from the same random field as the dataset5 from Test 1a. Remarkably is that RMSE for the dataset5 ($100.000$ locations) in Test 1a (5th row in Table~\ref{t:S1a_params}) is equal $0.0054$, whereas RMSE for similar dataset1 ($1.000.000$ locations) from Test 2b is equal $0.25$. This fact that the MLE approach is working better on a smaller sample,  can be explained by the fact that 1) the MLE approach (in general, even without $\H$-matrices) faces difficulties with large matrices, since the condition number of $\bC$ is increasing, and 2) the number of needed iterations in the MLE optimization procedure increases. We did not run RF and FCNN methods on Test 2b because the computing time is much larger than for the kNN time. The kNN time for Test 2b is 1.23 sec. 

\begin{figure}[h!]
\centering
\includegraphics[width=0.46\textwidth]{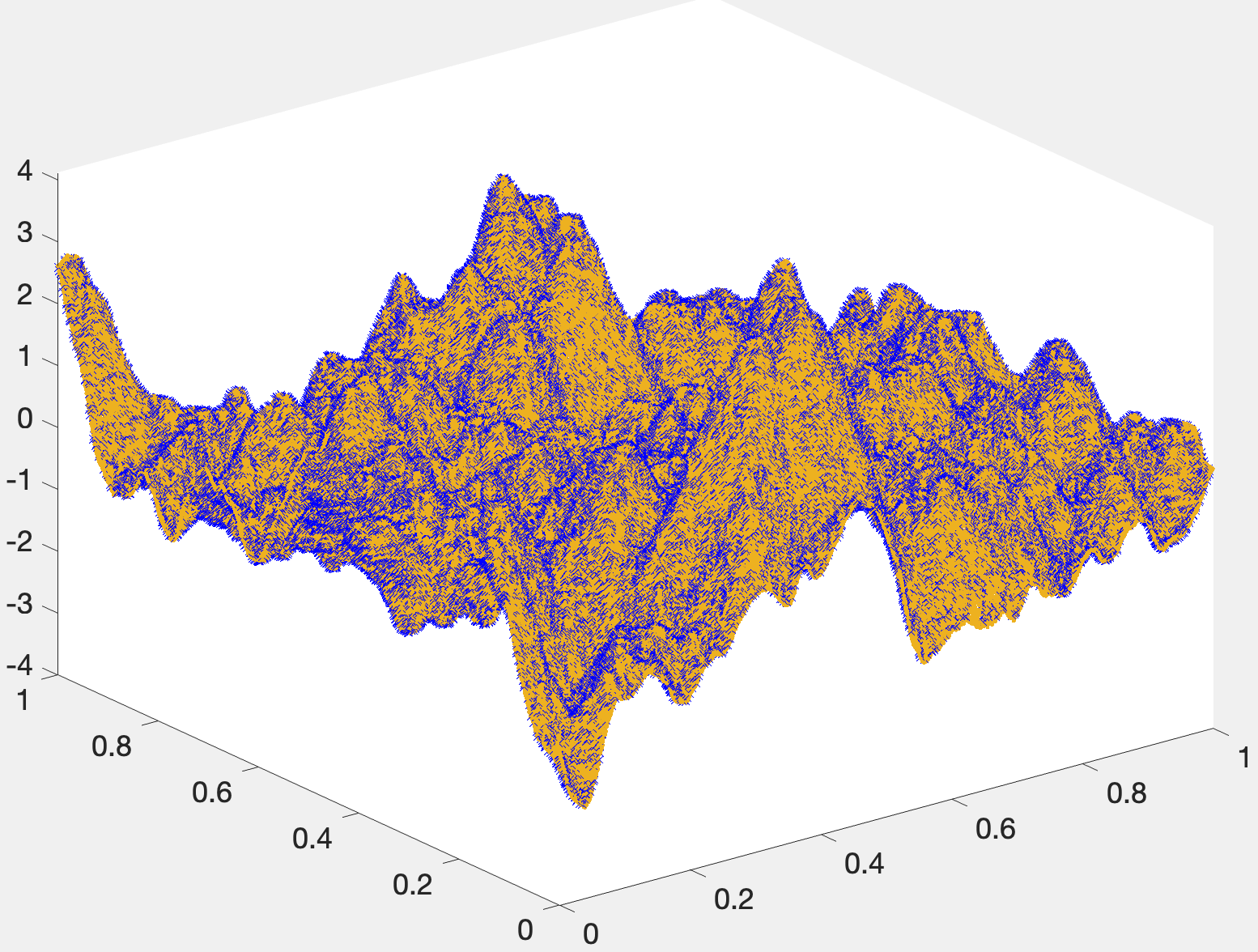}
\includegraphics[width=0.53\textwidth]{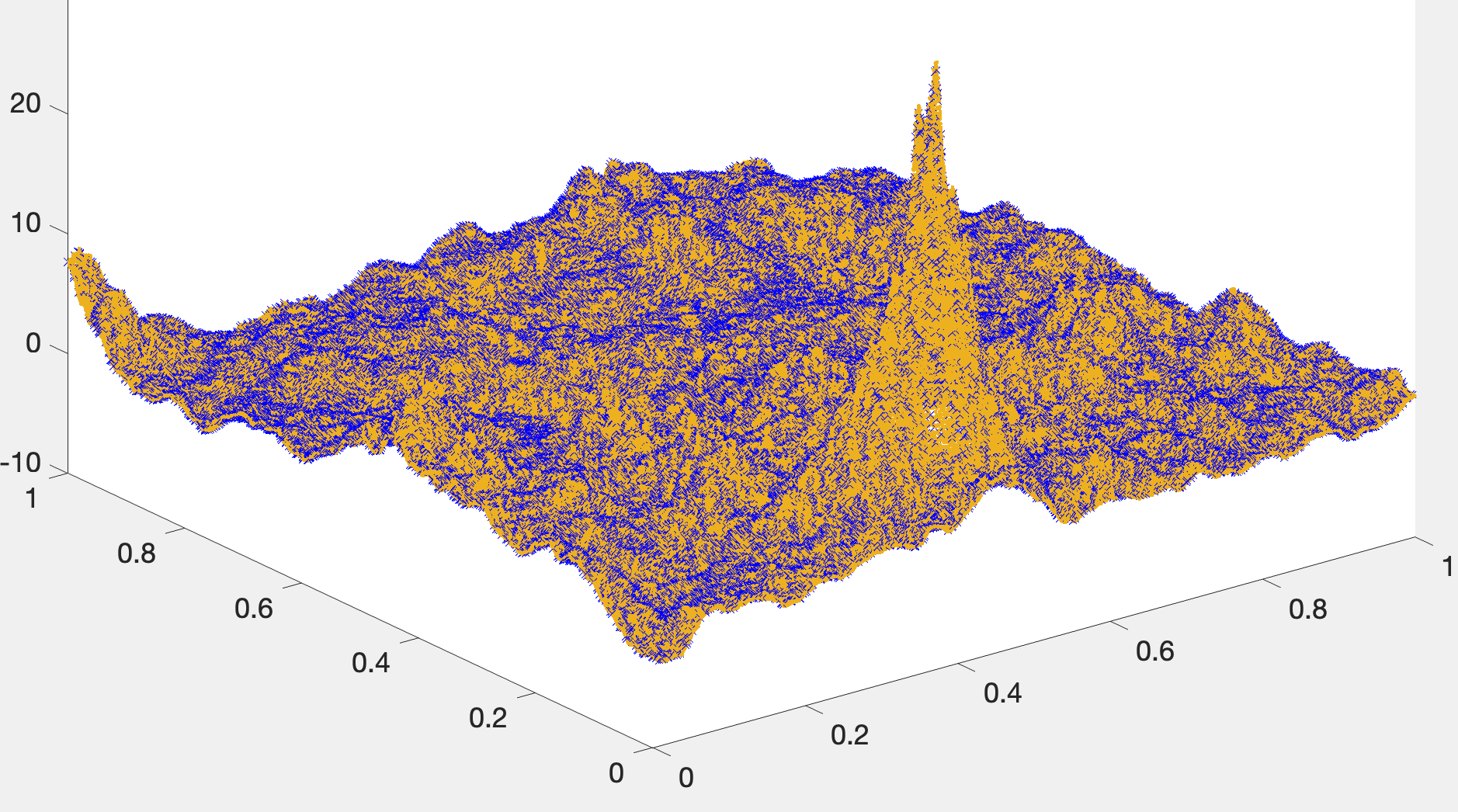}
\caption{Test 2b, datasets 1(left) and 2(right): Prediction obtained by the kNN method. The yellow points at $900.000$ locations were used for training and the blue points were predicted at $100.000$ new locations. One can see a very good alignment of both.}
\label{fig:Test2b-1_kNN}
\end{figure}

Predictions for datasets 1 and 2 from Test 2b are shown in Fig.~\ref{fig:Test2b-1_kNN}. The training datasets are depicted by yellow points and kNN predictions by blue points. We can see that the kNN method provides very good results.\\
\begin{table}[h!]
\centering
\begin{tabular}{|c|c|c|c|c|} \hline
&  \multicolumn{2}{c|}{Test 2a} &  \multicolumn{2}{c|}{Test 2b} \\ \hline
dataset & 1 & 2 & 1& 2 \\ \hline
$\H$-MLE         & 0.057  & 0.14 & 0.25 & 0.021  \\ \hline
kNN & 0.129 & 0.357 & 0.007 & 0.04  \\ \hline
RF     & 0.226 & 0.607 & --- & ---  \\ \hline
FCNN     & 0.243 & 0.74 & --- & ---  \\ \hline
\end{tabular}
\caption{Comparison of RMSE errors for $\H$-MLE, kNN, RF, and FCNN methods.}
\label{t:S2_RMSE_errors_kNN}
\end{table}
%
%

%
%

%
\newpage
\section{Conclusion}
We developed the $\H$-MLE procedure to estimate unknown parameters and to make statistical predictions at new locations.
In order to make computations faster, we approximated the joint Gaussian log-likelihood function in the $\H$-matrix format. In the numerical section, we considered $8+2+2=12$ datasets: 8 in Tests 1a and 1b, 2 in Test 2a, and 2 in Test 2b. All datasets in Tests 1a, 1b and 2a contain  
$90,000$ locations for training and $10,000$ for testing (prediction). Both datasets in Test 2b contained $900,000$ locations for training and $100,000$ for prediction.

The $\H$-matrix technique drastically reduces the required memory 
and computing time, making it possible to work with larger sets of observations obtained on unstructured meshes.
The main drawback of using $\H$-matrices is that too many linear algebra operations are required to estimate just four scalar unknown parameters.
For example, for some datasets with the unlikely chosen initial guess, we needed 400 iterations. On each iteration, we computed one $\H$-Cholesky factorization, one scalar product and solved a linear system. In total, it can take up to 8 hours on a modern parallel node for the dataset with $900, 000$ locations. A possible remedy is to precompute the initial guess. It will significantly reduce the required number of iterations and the total time. A good initial guess can be found, for instance, on a smaller subset of observations. Another drawback is that the $\H$-matrix approximation of $\bC$ and $\bL$ was recomputed entirely on every iteration for each new value of $\tau$ and $\sigma$. It would be a lot cheaper just to update already existing matrices by adding a new diagonal or by scaling $\bC$. It is indeed possible to modify the optimization algorithm, but the whole procedure will become more complicated. And the last drawback is that the total complexity depends on the matrix size and the number of parameters. For one to four parameters, the total computing time is acceptable, but it will be too large for five or more parameters. To tackle problems with large number of parameters we suggest to use low-rank tensor methods \cite{litv17Tensor,Grasedyck13, Khor_tensor_book18}. 

Among all implemented ML methods (kNN, random forest, deep neural network), the best results (for given datasets) were obtained by the kNN method with three or seven neighbors depending on the dataset. 
The results computed with the $\H$-MLE method were compared with the results obtained by the kNN method. 
For Test 2a, the $\H$-MLE method showed a smaller RMSE error than the kNN method, whereas, for Test 2b, the kNN method was better.
To conclude, it is not surprising that our $\H$-MLE method worked fine on most datasets. We also understand that we can improve the $\H$-MLE results simply by taking a smaller threshold and more accurate $\H$-matrix arithmetics. What surprised us is that the well-known and straightforward kNN method performed very good and very fast. Since we did not make any theoretical comparison and compared $\H$-MLE and ML methods only numerically on given datasets, we can not in general conclude which method is better. We also remind that we used kNN only for the prediction.

\textbf{Acknowledgment}
The study was carried out within the framework of the state contract of the Sobolev Institute of Mathematics (project no 0314-2019-0015). The work was partly supported by RFBR grant 19-29-01175. A. Litvinenko was supported by funding from the Alexander von Humboldt Foundation.

\bibliographystyle{plain}
\bibliography{Hcovariance.bib}
\end{document}